\documentclass[aps,preprint,superscriptaddress]{revtex4}

\usepackage{natbib}
\usepackage[dvips]{graphicx}
\usepackage{amsmath}
\usepackage{amssymb}

\begin{document}

\title{Two-Stage Bulk Electron Heating in the Diffusion Region of Anti-Parallel Symmetric Reconnection}

\author{A. Le}
\affiliation{Los Alamos National Laboratory, Los Alamos, New Mexico 87545, USA}
\author{J. Egedal}
\affiliation{University of Wisconsin---Madison, Madison, Wisconsin  53706, USA}
\author{W. Daughton}
\affiliation{Los Alamos National Laboratory, Los Alamos, New Mexico 87545, USA}

\begin{abstract}

Electron bulk energization in the diffusion region during anti-parallel symmetric reconnection entails two stages. First, the inflowing electrons are adiabatically trapped and energized by an ambipolar parallel electric field. Next, the electrons gain energy from the reconnection electric field as they undergo meandering motion. These collisionless mechanisms have been decribed previously, and they lead to highly-structured electron velocity distributions. Nevertheless, a simplified control-volume analysis gives estimates for how the net effective heating scales with the upstream plasma conditions in agreement with fully kinetic simulations and spacecraft observations.
\end{abstract}

\maketitle

\section{Introduction}
Magnetic reconnection is invoked to explain plasma heating and energization in a variety of space environments, including Earth's magnetosphere and the solar corona \cite{yamada:2010}. Despite kinetic numerical simulations and theoretical work \cite{hoshino:2001B,jaroschek:2004,loureiro:2013}, it remains unclear precisely how magnetic energy is converted to particle kinetic energy during collisionless reconnection and how the conversion process depends on the plasma conditions. Analysis of spacecraft data collected in Earth's magneto-tail \cite{eastwood:2013} and a laboratory reconnection experiment \cite{yamada:2015} suggest that $\sim 10-20\%$ of the magnetic energy released by reconnection is carried by the bulk electrons.

We develop a model for the electron heating through the diffusion region of symmetric, anti-parallel magnetic reconnection. The diffusion region is a small volume where electron kinetic effects break the frozen flux condition, and it is a main focus of NASA's MMS mission \cite{burch:2014}. While the electron heating in the diffusion region has been studied before \cite{hesse:1999,ricci:2003}, no previous model has offered predictions for how the electron temperature increase depends on the plasma conditions.

\begin{figure}[h]
\includegraphics[width = 10.0cm]{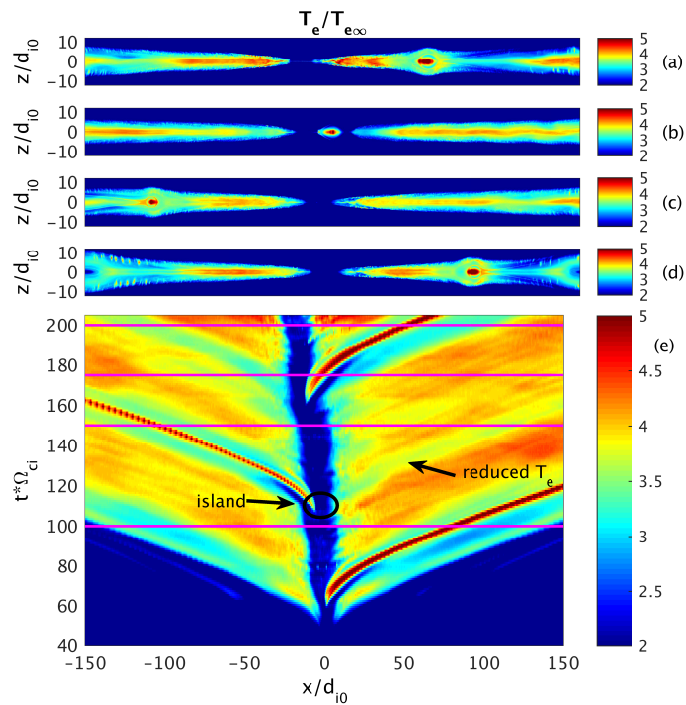}
\caption{Electron temperature profiles from an open-boundary PIC simulation of anti-parallel reconnection ($\beta_{e\infty}\sim0.04$, $m_i/m_e=50$, $T_i/T_e=5$, and $n_\infty/n_0=0.25$) at $t*\Omega_{ci}=$ (a) 200, (b) 175, (c) 150, and (d) 100. (e) The electron temperature along the mid-plane ($z=0$) is plotted over time, with magenta contours indicating the times selected for panels (a-d). Magnetic island formation in the diffusion region decreases the electron heating there, and $T_e$ remains decreased in the affected flux tubes $\gtrsim100$ $d_i$ downstream on both sides of the exhaust, not only the side containing the ejected island.  \label{fig:teopen}}
\end{figure}

In previous work, we found that the electrons are energized anisotropically as they advect towards the X-line in the reconnection inflow \cite{le:2010grl}. The fractional changes in parallel (with respect to the local magnetic field) and perpendicular temperatures turn out to scale in a predictable way with $\beta_{e\infty}$, the upstream ratio of electron to magnetic pressure. Here, we consider how the dependence on $\beta_{e\infty}$ extends into the electron energization within the diffusion region itself. The result is a model for the electron heating from upstream to the end of the diffusion region in which two main physical mechanisms operate: (1) the inflowing electrons are energized through an adiabatic trapping process \cite{egedal:2008jgr} and (2) the reconnection electric field then does additional work on a current sheet carried by meandering electrons within the diffusion region \cite{ng:2011,shuster:2015}. Relatively simple estimates developed here predict the additional heating within the diffusion region up to a factor of order unity that is determined empirically from kinetic simulations. The simulations cover a wide range of $\beta_{e\infty}$, and they demonstrate the limit of validity of the adiabatic model. 

The net increase in the electron temperature is found to scale roughly as $\Delta T_e/T_{e\infty} \propto 1/\beta_{e\infty}$. This scaling is consistent with a survey of THEMIS magnetopause reconnection observations \cite{phan:2013}. In the observations, the electron temperature increase $\Delta T_e$ from the inflow to the exhaust at least tens of ion inertial lengths downstream from the X-line scales with the upstream Alfven speed squared, $v_A^2$, which is equivalent to $\Delta T_e/T_{e\infty} \propto 1/\beta_{e\infty}$ for anti-parallel symmetric reconnection. Because only a small portion of the electrons in the far exhaust have passed directly through the diffusion region, it is not clear that the temperature scaling for the diffusion region and farther downstream should be the same. 

There is numerical evidence, however, that the diffusion region heating is related to the downstream temperature. In a numerical scaling study \cite{shay:2014}, the peak electron temperature was found to become nearly uniform from the end of the diffusion region out $\gtrsim 100$ $d_i$ into the exhaust. Additional evidence comes from the electron temperature illustrated in Figs ~\ref{fig:teopen}(a-d) at late times from a particle-in-cell (PIC) simulation of anti-parallel reconnection (similar to Ref.~\cite{le:2014}, but with an ion-to-electron temperature ratio of 5 and background density of 0.25 the peak Harris density). Figure~\ref{fig:teopen}(e) shows the time evolution of the electron temperature along the mid-plane ($z=0$). Note that the peak electron temperature is reached first at the end of the diffusion region near $x=0$, and this value then fills the exhaust as reconnected field lines are advected downstream at the outflow velocity. This suggests that additional energization processes \cite{dahlin:2014,egedal:2015} allow the elevated temperature set in the diffusion region to fill the exhaust as reconnected flux is carried downstream. Furthermore, the formation of magnetic islands within the diffusion region interrupts the diffusion region heating. The exhaust electron temperature in the affected flux tubes remains decreased as the reconnected plasma travels downstream. This modulation of the exhaust temperature is noticeable on both sides of the X-line, not only the side that contains the ejected island. Because reconnection is driven by steady inflow boundary conditions in this simulation, neither the global reconnection rate nor the overall geometry of the exhaust without an island changes. This provides further evidence that the initial heating in the diffusion region plays some role in setting the ultimate temperature of the downstream exhaust. 

In Section~\ref{sec:pic}, we describe the PIC simulations we use to guide and confirm our model of electron heating. The kinetic energization processes and their resulting electron velocity distributions are presented in Section~\ref{sec:pdf}. Some details of a previous inflow heating (Section~\ref{sec:inflow}) model and a new simplified diffusion region heating model (Section~\ref{sec:edr}) follow.

\section{Particle-in-Cell Simulations}
\label{sec:pic}
To study electron energization in the diffusion region, we performed a series of 2D PIC simulations of anti-parallel Harris sheet reconnection using the code VPIC \cite{bowers:2008} with open boundary conditions \cite{daughton:2006}. In our coordinates, the initial magnetic field and density are $B_x = B_0\tanh(z/\lambda)$ and $n = n_0\cosh^2(z/\lambda) + n_\infty$, where $\lambda = 0.5 d_{i0} = 0.5 \sqrt{\epsilon_0 m_ic^2/n_0 e^2}$. The computational domain is $4096\times4096$ cells $=50d_{ip}\times50d_{ip}$, where $d_{ip}$ is the ion inertial length based on the peak density $n_p = n_0 + n_\infty$. Other parameters are initial uniform temperatures with $T_{i\infty}/T_{e\infty} = 5$, $\omega_{pe0}/\omega_{ce0} = 2$, and $\sim400$ particles per species per cell. We use a mass ratio of $m_i/m_e = 200$ because a ratio of $m_i/m_e\gtrsim100$ is necessary to capture the dynamics of electrons that follow adiabatic trapped orbits \cite{le:2009}. The background density $n_\infty$ is varied so the initial upstream electron beta is $\beta_{e\infty} = 2\mu_0 n_\infty T_{e\infty}/B_0^2 = 2^k$ with $k = -7,-6,...,-1$. 

Besides the main scan with varying $n_\infty$, we test the dependence of the electron heating on upstream electron temperature $T_{e\infty}$ with two additional runs performed at fixed $\beta_{e\infty} = 1/8$, with $n_\infty/n_0 = 3/2$ for $T_{e\infty}/T_{eH} = 1/2$ and $n_\infty/n_0 = 3/8$ for $T_{e\infty}/T_{eH} = 2$ (where $T_{eH}$ is the Harris sheet electron temperature). To study the dependence on mass ratio, we use three runs reported on previously \cite{le:2013} with varying mass ratio $m_i/m_e$. Finally, while an exhaustive parameter scan is not feasible in 3D, data from the 3D run of \cite{chen:2012} (with $m_i/m_e=300$ and $\beta_{e\infty}=0.05$) is presented and is consistent with the model. Indeed, 3D effects \cite{che:2010} have been found not to substantially alter the dynamics of anti-parallel reconnection in typical magnetospheric regimes with $T_i>T_e$ \cite{roytershteyn:2012}.

\begin{figure}[h]
\includegraphics[width = 8.0cm]{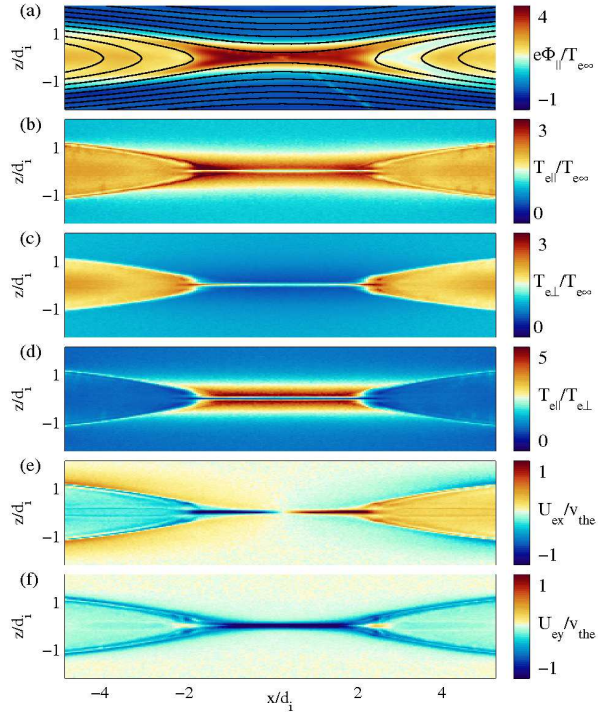}
\caption{Profiles from a PIC run with $\beta_{e\infty}=1/16$. (a) The pseudo-potential $e\Phi_{\parallel}/T_{e\infty}$ with sample in-plane field lines. The (b) parallel and (c) perpendicular electron temperatures. (d) The electron temperature anisotropy $T_{e||}/T_{e\perp}$ peaks near 5 in the inflow. The (e) outflow (x-directed) and (f) out-of-plane (y-directed) electron flow velocities. \label{fig:e-4}}
\end{figure}

Each simulation was run until reconnection reached a quasi-steady state. Typical field profiles during quasi-steady reconnection are plotted in Fig.~\ref{fig:e-4} from the run with $\beta_{e\infty}=1/16$. As described previously \cite{egedal:2013pop}, a parallel electric field structure develops in the inflow to trap electrons and maintain quasi-neutrality. The effect of electric trapping is measured by the pseudo-potential $\Phi_{\parallel} = \int E_\parallel dl$ (the integral of the parallel electric field along magnetic field lines). $\Phi_{\parallel}$ in Fig.~\ref{fig:e-4}(a) peaks at $e\Phi_{\parallel}\sim 4T_{e\infty}$ and traps the bulk electrons. Combined with the perpendicular cooling that results from $\mu$ conservation, the trapping results in a temperature anisotropy with  $T_{e\parallel}>T_{e\perp}$ [Figs.~\ref{fig:e-4}(b) and (c)] that obeys known equations of state \cite{le:2009}, and it reaches $T_{e\parallel}/T_{e\perp}\sim 5$ [Fig.~\ref{fig:e-4}(d)]. The temperature anisotropy supports the current layer near the X-line [Figs.~\ref{fig:e-4}(e) and (f)], which is peaked in a sheet of width 2---4$d_{e}$ and length $50d_{e}$ ($d_{e}$ based on the density $2d_i$ upstream of the X-line). We refer to this area as the electron diffusion region.

\section{Velocity Distributions and Heating Mechanisms}
\label{sec:pdf}

\begin{figure*}
\includegraphics[width = 14cm]{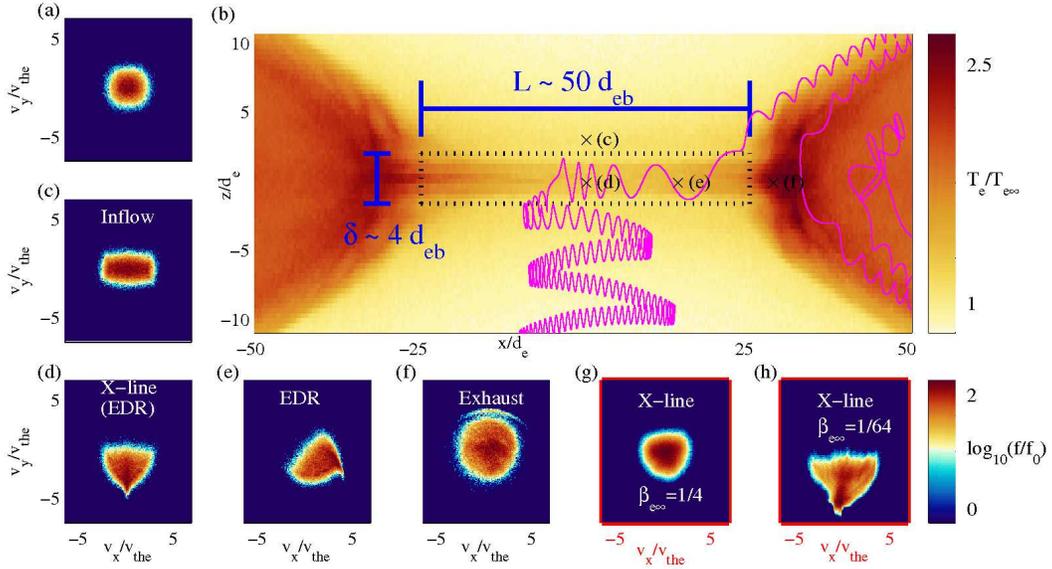}
\caption{(b) Our model is based on electron energy balance over the dotted box, referred to here as the electron diffusion region. Typical electron orbits (like the sample one shown in magenta) undergo meandering motion in the electron diffusion region. (a) The reduced electron velocity distribution $f(v_x,v_y)$ $15d_e$ upstream from the X-line is Maxwellian. (c)-(f) Velocity distributions at the points labeled by $\times$'s in (b). Electron distributions at the X-line in simulations with (g) higher and (h) lower $\beta_{e\infty}$ show the tail of electrons accelerated to large $|v_y|$ is larger in electron distributions at low $\beta_{e\infty}$. \label{fig:box}}
\end{figure*}

The electron temperature $T_e = (T_\parallel + 2T_\perp)/3$ is plotted in Fig.~\ref{fig:box}(b) with the boundary of the electron diffusion region indicated by a dotted box. A typical electron orbit is also shown. The electron follows a trapped trajectory in the inflow, characterized by the repeated reversal of its parallel ($\sim x$) velocity direction.  Sample reduced electron velocity distributions $f(v_x,v_y)$ are plotted in Figs.~\ref{fig:box}(c-f) at each point marked by an $\times$ in Fig.~\ref{fig:box}(b). The upstream distribution [Fig.~\ref{fig:box}(a)] is Maxwellian. The inflow distribution [Fig.~\ref{fig:box}(c)] is a typical trapped distribution that is elongated in the parallel ($\sim x$) direction \cite{egedal:2013pop}. 

The electron orbit in Fig.~\ref{fig:box}(b) then undergoes meandering motion \cite{speiser:1965} within the diffusion region and oscillates across the inner current layer. Here within the diffusion region itself, the velocity distributions more finely structured. As electrons meander across the central sheet, they gain energy from the reconnection electric field $E_y$. This process produces a distribution with a striated triangular tail extended in $v_y$ near the X-line [as in Figs.~\ref{fig:box}(d)] \cite{ng:2011}, where each striation is composed of electrons that bounced a set number times across the diffusion region layer. Further downstream, the reconnected magnetic field component $B_z$ mixes the striations [Fig.~\ref{fig:box}(e)] \cite{bessho:2014}. Finally, in the exhaust [Fig.~\ref{fig:box}(f)], the typical electron orbits include regions of chaotic motion, and pitch angle mixing produces nearly isotropic velocity distributions \cite{buchner:1989,le:2010grl,chen:2008}.

As far as bulk energization, the electron distributions thus indicate two separate processes. First, the inflow trapping elongates the velocity distributions and increases the effective parallel temperature by $\Delta T_{\parallel}$ while adiabatic cooling from $\mu$ conservation leads to a decrease $\Delta T_\perp$. We denote the initial total temperature increment $\Delta T_{{inflow}} = T_{{in}} -T_{e\infty}=(\Delta T_{\parallel}+2\Delta T_{\perp})/3$. Next, the meandering electrons within the diffusion region are accelerated to larger $|v_y|$ by the reconnection electric field, forming the tips of the triangular velocity distributions. Chaotic mixing finally produces isotropic distributions in the exhaust with an increased effective temperature $T_{{out}}$. We refer to this additional temperature increase within the diffusion region as $\Delta T_{{diff}}=T_{{out}}-T_{{in}}$. Both $\Delta T_{{inflow}}$ and $\Delta T_{{diff}}$ depend on the upstream plasma conditions. The differences at varying $\beta_{e\infty}$ are evident in the velocity distributions computed near the X-line in Figs.~\ref{fig:box}(g) and (h). The distributions are only mildly anisotropic and have a small tail accelerated by $E_y$ at $\beta_{e\infty}=1/4$, while there is strong anisotropy and a tail accelerated to large $|v_{ey}|$ for $\beta_{e\infty}=1/64$.

\section{Inflow Heating}
\label{sec:inflow}

We now review a model for the inflow heating $\Delta T_{{inflow}}$ and develop estimates for the diffusion region heating $\Delta T_{{diff}}$ to determine how the temperature increments scale with the plasma parameters. Our results for the inflow heating derive from the fact that the electron temperature anisotropy and the pseudo-potential $\Phi_{\parallel}$ saturate when the electrons approach the firehose instability threshold, $F\equiv\mu_0(p_{e\parallel}-p_{e\perp})/{B^2}\simeq 1$. This follows from gross momentum balance, which requires electron anisotropy to balance a large fraction of the $\bf{J\times B}$ magnetic "tension" force on the electron jets that flow in the diffusion region. Typically, the electron temperature anisotropy therefore peaks at a value $F\lesssim 1$ immediately upstream from the electron jets \cite{le:2010grl,egedal:2013pop}.

\begin{figure}
\includegraphics[width = 8cm]{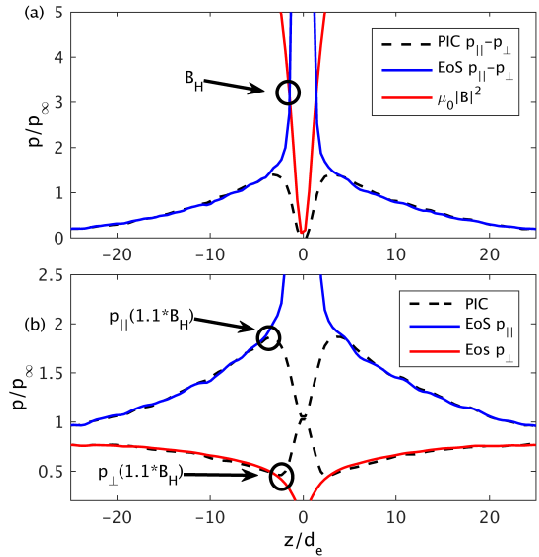}
\caption{Cuts in $z$ across the electron diffusion region. (a) The magnetic field upstream of the diffusion region $B_H$ is found by solving the firehose condition $(p_{e\parallel}-p_{e\perp})\simeq\mu_0B_H^2$ based on equations of state \cite{le:2009} for $p_{e\parallel}(n,B)$ and $p_{e\perp}(n,B)$. (b) The pressure components at peak anisotropy may then be found by evaluating the equations of state at $B=1.1B_H$. \label{fig:bheos}}
\end{figure}

We follow the method of \cite{le:2010grl} to estimate the inflow anisotropic heating. First, upstream parameters $n_\infty$, $B_\infty$, and $T_{e\infty}$ are found $2 d_i$ upstream from the X-line to use in normalizing the equations of state for the inflow electrons. These may differ by up to $~25\%$ from the initial asymptotic values. Next, we numerically solve $(p_{e\parallel}-p_{e\perp})/{B^2}=1$ based on the equations of state for adiabatically trapped electrons, $p_{e\parallel}(n/n_{\infty},B/B_\infty)$ and $p_{e\perp}(n/n_{\infty},B/B_\infty)$ \cite{le:2009}, to determine the magnetic field strength $B_H$ upstream from the diffusion region, measured where the electron current falls to $40\%$ of its peak value. This is illustrated in Fig.~\ref{fig:bheos}(a). Note that for anti-parallel reconnection, the equations of state do not apply in the center of the diffusion region where the electron orbits are unmagnetized. Although they typically break down before the value $B_H$ is reached, the equations of state at the firehose threshold give a good estimate for the magnetic field immediately outside the current layer. Finally, it is found empirically that evaluating the equations of state with $B=1.1 B_H$ (the factor of 1.1 provides a better fit over the larger range of $\beta_{e\infty}$ studied here than the factor of 1.25 used originally in \cite{le:2010grl}) and $n=n_\infty$ yields good estimates for both the peak pseudo-potential $\Phi_{\parallel}$ and the electron temperatures $T_{e\parallel}$ and $T_{e\perp}$. See Fig.~\ref{fig:bheos}(b). 

\begin{figure}
\includegraphics[width = 8cm]{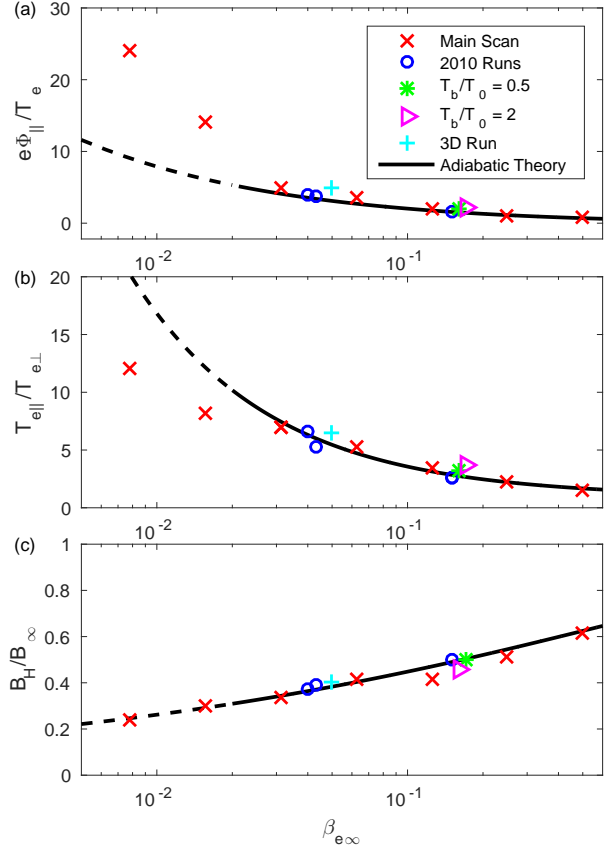}
\caption{The (a) trapping potential $e\Phi_{\parallel}/T_{e\infty}$, (b) peak temperature anisotropy $T_{e\parallel}/T_{e\perp}$, and (c) magnetic field strength $B_H$ upstream from the diffusion region as functions of upstream $\beta_{e\infty}$. A limited range (three circles) was studied previously \cite{le:2010grl,le:2010pop}. Two additional runs at $\beta_{e\infty}=0.125$ with varying $T_{e\infty}$ and the 3D run of Ref.~\cite{chen:2012} are also included. At very low $\beta_{e\infty}<0.03$, the adiabatic assumption breaks down, and the effective heating is larger than predicted by the adiabatic model \cite{egedal:2015}. \label{fig:curves}}
\end{figure}

The model of \cite{le:2010grl} was confirmed numerically for a somewhat limited range of $\beta_{e\infty}$. Meanwhile, additional simulations \cite{egedal:2015} showed that the adiabatic model breaks down at low $\beta_{e\infty}< 0.03$. The range of $\beta_{e\infty}$ considered here covers the span of parameters from low $\beta_{e\infty}$ where the adiabatic model begins to fail to high $\beta_{e\infty}$ where the weak inflow heating is very difficult to resolve both in spacecraft data and in particle-in-cell simulations. Predictions and simulation data are plotted in Fig.~\ref{fig:curves}. For low $\beta_{e\infty}$, the adiabatic model outlined above predicts the pseudo-potential scales as $e\Phi_{\parallel}/T_{e\infty}\propto \beta_{e\infty}^{-1/2}$, the peak temperature anisotropy scales as $T_{e\parallel}/T_{e\perp}\propto \beta_{e\infty}^{-3/4} $, and $B_H \propto \beta_{e\infty}^{1/4}$. Our scan of kinetic runs demonstrates that the parallel pseudo-potential in the inflow becomes larger than predicted by the adiabatic model at very low $\beta_{e\infty}$. The peak temperature anisotropy $T_{e\parallel}/T_{e\perp}$ decreases somewhat. This results, however, from an increase in $T_{e\perp}$ above the adiabatic predictions. 
The net energization $(\Delta T_{e\parallel}+2\Delta T_{e\perp})/3$ is greater than predicted by the adiabatic model.

\section{Diffusion Region Heating}
\label{sec:edr}

\begin{figure}
\includegraphics[width = 8cm]{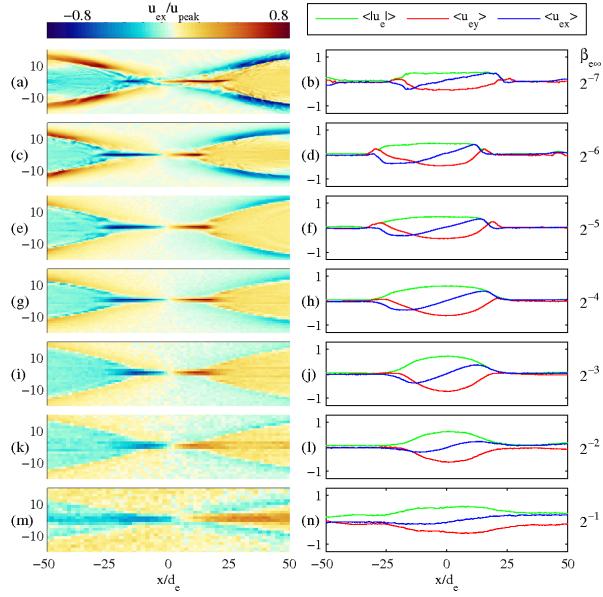}
\caption{Left-hand panels: electron outflow velocity $u_{ex}$ profiles normalized to peak $u_{ey}$ at the X-line. Right-hand panels: mean electron flow profiles averaged over the width of the diffusion region.\label{fig:uex}}
\end{figure}

\begin{figure}
\includegraphics[width = 8cm]{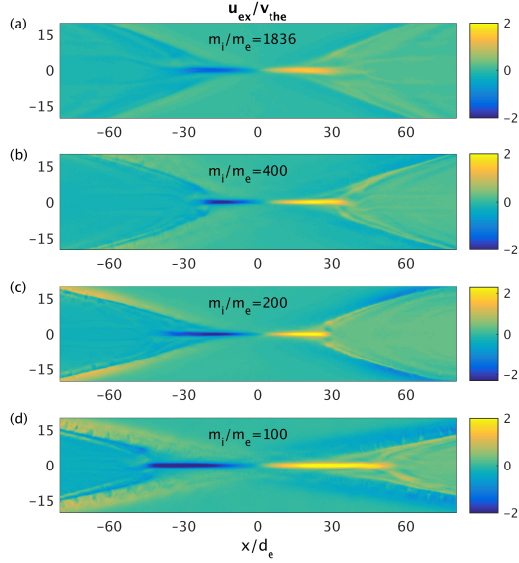}
\caption{Electron outflow velocity $u_{ex}$ profiles at varying mass ratio $m_i/m_e$.\label{fig:uexmime}}
\end{figure}

A simplified model is derived below that reproduces the scalings with plasma parameters we observe in a large number of kinetic simulations, even though it uses several approximations concerning the morphology of the electron diffusion region current layers. A single overall coefficient of order unity is determined empirically to fit the model to the simulation data. To estimate the diffusion region heating, we consider the electron conservation equations for mass, momentum, and energy over a control volume that covers the diffusion region of length $L\simeq 50d_{e}$ and width $\delta\sim$ 2---4$d_e$. These typical length scales cover the electron jets in the diffusion region, as plotted in Fig.~\ref{fig:uex} where the length scales are normalized to $d_e$ based on the upstream density. The dependence on mass ratio $m_i/m_e$ is plotted in Fig.~\ref{fig:uexmime}. While a previous study found that the size of the diffusion region scales as $\beta_{e\infty}^{1/4}$ \cite{nakamura:2016}, our runs did not confirm this scaling. The precise length of the diffusion region may be sensitive to boundary conditions and the time selected for measuring its length. In any case, the nominal sizes we choose are sufficient for our estimates within the range of parameters relevant to magnetospheric reconnection that we consider.

Number or mass conservation demands that the inflowing flux of electrons $\sim nu_{ez}L$ and the diffusion region exhaust flux $\sim nu_{ex}\delta$ be equal, and we denote this single particle flux $\Gamma_e$. Furthermore, a current sheet supported by anisotropy consistent with the equations of state for the inflowing electrons also requires that the magnetic field strength just upstream from the electron jet be uniform and equal to the value $B_H$. By Ampere's law, this magnetic field is proportional to the net current in the sheet. While maintaining its magnitude $B_H$, the magnetic field rotates in direction with the electron flow \cite{hesse:2008}. As shown previously \cite{le:2010pop}, the peak net out-of-plane flux within the diffusion region is therefore also of the same order as the in-plane flux, $\Gamma_e \simeq nu_{ey}\delta$. The left panels of Fig.~\ref{fig:uex} show the electron outflow jets, and the right-hand panels show mean electron flow profiles $<u_{ex}>$, $<u_{ez}>$, and $|u_e|$ averaged over the width of the diffusion region. At higher $\beta_{e\infty}$, the peak $u_{ex}$ does not  reach the peak $u_{ey}$ \cite{divin:2012}. This approximation concerning the magnitude of the electron fluxes is thus roughest in the high $\beta_{e\infty}$ cases where the electron jets do not flow much faster than the background Alfvenic outflow. In these cases, the diffusion region electron heating is small, and it is at the limit of resolution in both PIC simulations and spacecraft observations.

As confirmed by the kinetic simulations, the main contributions to energy balance integrated over the diffusion region are advection across the boundaries and work done through $\bf{E\cdot u_e}$. An exception is at the lowest $\beta_{e\infty}$ of $\sim0.008$, where the thermal heat flux driven by instabilities carries nearly $\sim 10\%$ of the energy dissipated by $\bf{E\cdot u_e}$. The electric field does work on the electrons (per unit length in the out-of-plane direction) of size 
\begin{equation}
\int_{{box}}ne{\bf{E\cdot u_e}}dA \sim \frac{1}{2}e E_{{rec}} n u_{ey} \delta L \sim {\frac{1}{2}}e \Gamma_e E_{{rec}} L,
\label{eq:edotj}
\end{equation}
where the factor of $1/2$ accounts for the current density profile falling off from its peak value at the center of the sheet. For our estimates, we take a typical value for the reconnection electric field of $E_{{rec}}\simeq 0.1v_{A\infty} B_{\infty}$, although $E_{rec}$ tends to decrease with increasing upstream density \cite{wu:2011}. Meanwhile, the net electron thermal energy advected out of the box is
\begin{eqnarray}
\oint_{{box}} {\bf{u}}_e\cdot(\frac{3}{2}p_e{I} + {P}_e) \cdot {\bf\hat{n}} dl &\sim& \frac{5}{2} (nu_{ex}T_{{out}}2\delta-nu_{ez}T_{{in}}2L) \nonumber \\
 &\sim& 5\Gamma_e\Delta T_{{diff}},
\label{eq:pu}
\end{eqnarray}
Balancing the contributions from Eqs.~\ref{eq:edotj} and \ref{eq:pu}, the temperature change across the diffusion region is $\Delta T_{{diff}}\sim e E_{{rec}} L/10$, or in terms of the upstream conditions,
\begin{equation}
\frac{\Delta T_{{diff}}}{T_{e\infty}} \simeq \frac{C*e(0.1v_{A\infty} B_{{\infty}})(50 d_e)}{10T_{e\infty}} \simeq \frac{C}{\beta_{e\infty}}\sqrt{\frac{m_e}{m_i}}.
\label{eq:Test}
\end{equation}
Our approximations, which include neglecting the precise current and field profiles, turn out to overestimate the diffusion region heating. We thus introduce the factor of $C=0.55$ to fit the simulation data.

\begin{figure}
\includegraphics[width = 14cm]{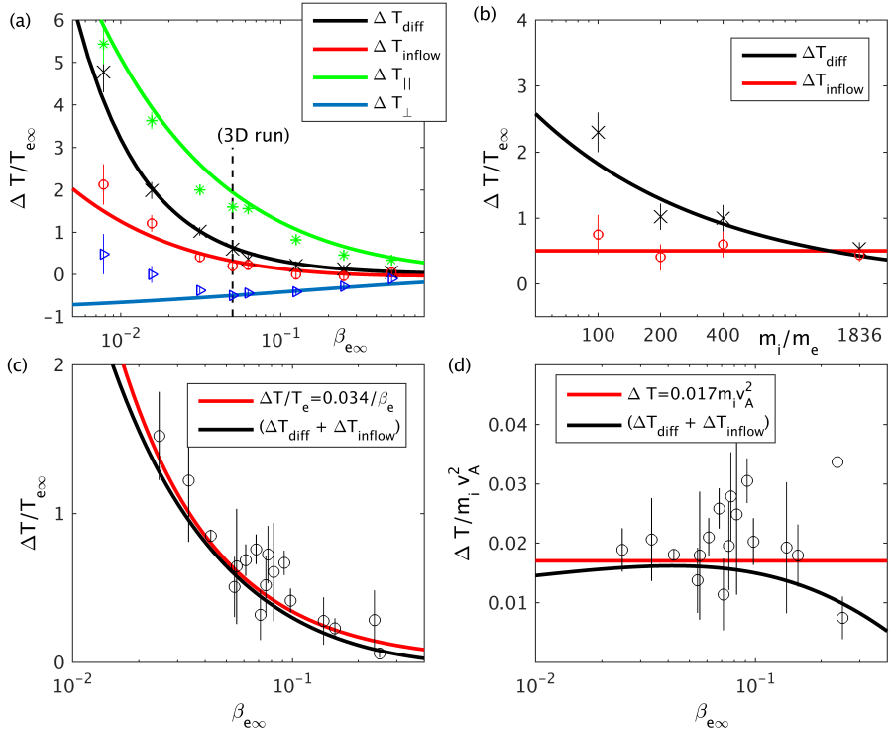}
\caption{(a) Changes in electron temperature depend on the upstream $\beta_{e\infty}$. The 3D run of \cite{chen:2012} had $m_i/m_e=300$ rather than 200. (b) $\Delta T_{{inflow}}$ and $\Delta T_{{diff}}$ as functions of $m_i/m_e$ at fixed $\beta_{e\infty}\simeq0.03$ (cases for $m_i/m_e = 100$, 400, and 1836 from \cite{le:2013}). (c) The model predictions agree reasonably well with the empirical scaling $\Delta T_e/T_{e\infty}\simeq 0.034/\beta_{e\infty}$ and data from 17 reconnection events with magnetic shear angle $>165^{\circ}$ observed by THEMIS \cite{phan:2013}. (d) Same data as (c) plotted with temperature increments normalized to $m_i v_A^2$. \label{fig:delta}}
\end{figure}

The model predictions are plotted in Fig.~\ref{fig:delta}(a). The model for first stage of heating, involving adiabatic trapping in the inflow, predicts both $\Delta T_{\parallel}$ and $\Delta T_{\perp}$, and the predictions are plotted along with the total $\Delta T_{{inflow}}$. Contrary to predictions, $T_{\perp}$ increases somewhat for $\beta_{e\infty}< 0.03$. This results from non-adiabatic motion and streaming instabilities, and it highlights the need for an extended model at low $\beta_{e\infty}$. Also plotted is the model for the second stage of heating within the diffusion region $\Delta T_{{diff}}$ based on Eq.~\ref{eq:Test}. Note that $\Delta T_{{diff}}$ in Eq.~\ref{eq:Test} scales as $(m_e/m_i)^{1/2}$ [see Fig.~\ref{fig:delta}(c) for numerical results]. On the other hand, $\Delta T_{{inflow}}$ is independent of the mass ratio as long as the trapped electron bouncing motion is adiabatic (usually for $m_i/m_e>~100$). The net temperature change $\Delta T_{e} = \Delta T_{{inflow}} + \Delta T_{{diff}}$ therefore scales more weakly than $(m_e/m_i)^{1/2}$ (similar to \cite{shay:2014}), although our model predicts that it in fact does not follow any simple power law in $(m_i/m_e)$. The results highlight that care must be taken when comparing various heating processes in kinetic simulations to observational data because different processes depend on the numerical parameters, such as mass ratio, in different ways. At the physical mass ratio of $m_i/m_e$=1836, the diffusion region heating of our model will typically be of a similar magnitude or smaller than the inflow heating.

Evaluating the model for the total temperature jump $\Delta T_{e}$ at the physical proton mass ratio yields the black curve Fig.~\ref{fig:delta}(c). It compares favorably to, though it is slightly lower than, the empirical scaling $\Delta T_e/T_{e\infty}\simeq0.034/\beta_{e\infty}$ (red curve) determined from the THEMIS survey of reconnection exhausts in magnetopause crossings \cite{phan:2013}. Observational data points from events with a weak guide field, corresponding to magnetic shear angles $>165^\circ$, are also plotted. The empirical scaling is equivalent to $\Delta T_e\simeq0.017 m_i v_{A\infty}^2$ for weak guide fields, and the data are re-plotted normalized to $m_i v_A^2$ in Fig.~\ref{fig:delta}(d). The model is consistent with the scaling $\Delta T_e\propto m_i v_{A\infty}^2$ to a good approximation for $\beta_{e\infty}\lesssim 0.25$. Again, it is somewhat unexpected that our model based on the diffusion region agrees fairly well with spacecraft data collected possibly $\sim100$ $d_i$ \cite{phan:pc} downstream from the X-line. This is consistent, however, with numerical calculations that show a uniform electron exhaust temperature \cite{shay:2014} and the relationship between the diffusion region dynamics and the downstream temperature presented in the Introduction. In the run of Fig.~\ref{fig:teopen}, islands within the diffusion region temporarily reduce the length of the electron current sheet by as much as 50\%, and as a result $\Delta T_{diff}$ is reduced from $\sim2T_{e\infty}$ (in agreement with Eq.~\ref{eq:Test}) to $\sim1$---$1.5T_{e\infty}$. The dip in temperature persists more than 100 $d_i$ downstream into the exhaust.

The model for electron heating has a couple of implications for high-resolution electron measurements such as those available from the MMS mission. First, it could possibly be combined with other methods, for example analyzing the striated diffusion region electron distributions \cite{bessho:2014}, to help place constraints on the size of $E_{rec}$ and the reconnection rate. Furthermore, the model suggests that the diffusion region striated electron distributions should be most apparent in low $\beta_{e\infty}$ reconnection. In this regime, $\Delta T_{diff}$ may be larger than the upstream $T_{\perp}$, and the striations should therefore be distinctly separated \cite{bessho:2014}. 

\section{Discussion and Summary}

In summary, a model for the diffusion region in symmetric anti-parallel reconnection predicts the electron heating. The first stage of energization occurs in the inflow because of adiabatic trapping, and it results in anisotropic heating with $\Delta T_{e\parallel}$ and $\Delta T_{e\perp}$ described by known equations of state. As demonstrated by our simulations spanning a wide range of $\beta_{e\infty}$, an extended model will have to be formulated for low $\beta_{e\infty}< 0.03$ when the adiabatic assumption breaks down. In the next stage, meandering electrons in the diffusion region are further energized by the reconnection electric field, and the temperature increases by an additional amount $\Delta T_{{diff}}$. It is worth noting that, unlike the adiabatic inflow heating for $m_i/m_e\>\sim100$, the diffusion region heating is sensitive to the mass ratio employed in numerical simulations. We developed a simple scaling law for this additional heating $\Delta T_{diff}$, and the composite two-stage model agrees with an empirical scaling based on spacecraft measurements taken downstream in the reconnection exhaust. While it was not evident that the diffusion region heating could affect the temperature far downstream, this dependence is consistent with simulations that show a flux tube retains the temperature set in the diffusion region as it is advected far into the exhaust.

It is interesting to note that while the individual anisotropic increases $\Delta T_\parallel$ and $\Delta T_\perp$ of the inflow depend sensitively on $T_{e\infty}$, the predicted total heating $\Delta T_{e}$ turns out to be roughly proportional to $v_{A\infty}^2$ to a good approximation for $\beta_{e\infty} < 0.25$. This scaling also held in the magnetopause observations \cite{phan:2013} and a previous numerical scaling study \cite{shay:2014}. A fluid model based on magnetized electrons appropriate for reconnection with a guide field also resulted in a nearly identical scaling for the electron temperature in the exhaust \cite{ohia:2015}. In the fluid model, the electron heating is related to overall pressure balance of the reconnection exhaust. For guide fields that are not too large, force balance requires the plasma pressure in the current sheet to compensate some significant fraction of the upstream magnetic pressure $\propto m_iv_{A\infty}^2$. It is thus natural that the total electron heating scales with the upstream magnetic pressure. The predicted level of heating and partition of energy between electrons and ions, however, depends in addition on the assumed equations of state for each species. Adiabatic ion and double adiabatic electron closures yielded good results, suggesting heat transport is somehow limited.

The observational survey \cite{phan:2013} and numerical study \cite{shay:2014} included a range of guide magnetic fields and plasma parameters. Meanwhile, it has been found that the mechanisms that energize the electrons in the far exhaust--such as parallel electric field acceleration and drifts in perpendicular fields \cite{egedal:2015}--play relatively smaller or larger roles depending on the strength of the guide field \cite{dahlin:2014}. It thus remains an open question how various electron processes, from adiabatic electric trapping to meandering diffusion region orbits to Fermi bounce acceleration in the far exhaust, conspire to yield relatively uniform electron temperature profiles that obey a simple scaling law to a good approximation for a broad range of plasma parameters found in the magnetosphere.

\begin{acknowledgments}
A.L. was supported by NASA grant NNX14AL38G at the Space Science Institute and by the LDRD office at LANL. J.E. acknowledges support through NSF GEM award 1405166 and NASA grant NNX14AC68G. W.D.'s work was supported by NASA's Heliophysics Theory Program. Simulations were performed on Pleiades provided by NASA's HEC Program and with LANL Institutional Computing resources. 
\end{acknowledgments}


\begin{thebibliography}{37}
\expandafter\ifx\csname natexlab\endcsname\relax\def\natexlab#1{#1}\fi
\expandafter\ifx\csname bibnamefont\endcsname\relax
  \def\bibnamefont#1{#1}\fi
\expandafter\ifx\csname bibfnamefont\endcsname\relax
  \def\bibfnamefont#1{#1}\fi
\expandafter\ifx\csname citenamefont\endcsname\relax
  \def\citenamefont#1{#1}\fi
\expandafter\ifx\csname url\endcsname\relax
  \def\url#1{\texttt{#1}}\fi
\expandafter\ifx\csname urlprefix\endcsname\relax\def\urlprefix{URL }\fi
\providecommand{\bibinfo}[2]{#2}
\providecommand{\eprint}[2][]{\url{#2}}

\bibitem[{\citenamefont{Yamada et~al.}(2010)\citenamefont{Yamada, Kulsrud, and
  Ji}}]{yamada:2010}
\bibinfo{author}{\bibfnamefont{M.}~\bibnamefont{Yamada}},
  \bibinfo{author}{\bibfnamefont{R.}~\bibnamefont{Kulsrud}}, \bibnamefont{and}
  \bibinfo{author}{\bibfnamefont{H.}~\bibnamefont{Ji}}, \bibinfo{journal}{Rev.
  Mod. Phys.} \textbf{\bibinfo{volume}{82}}, \bibinfo{pages}{603}
  (\bibinfo{year}{2010}).

\bibitem[{\citenamefont{Hoshino et~al.}(2001)\citenamefont{Hoshino, Hiraide,
  and Mukai}}]{hoshino:2001B}
\bibinfo{author}{\bibfnamefont{M.}~\bibnamefont{Hoshino}},
  \bibinfo{author}{\bibfnamefont{K.}~\bibnamefont{Hiraide}}, \bibnamefont{and}
  \bibinfo{author}{\bibfnamefont{T.}~\bibnamefont{Mukai}},
  \bibinfo{journal}{Earth, planets and space} \textbf{\bibinfo{volume}{53}},
  \bibinfo{pages}{627} (\bibinfo{year}{2001}).

\bibitem[{\citenamefont{Jaroschek et~al.}(2004)\citenamefont{Jaroschek,
  Treumann, Lesch, and Scholer}}]{jaroschek:2004}
\bibinfo{author}{\bibfnamefont{C.~H.} \bibnamefont{Jaroschek}},
  \bibinfo{author}{\bibfnamefont{R.~A.} \bibnamefont{Treumann}},
  \bibinfo{author}{\bibfnamefont{H.}~\bibnamefont{Lesch}}, \bibnamefont{and}
  \bibinfo{author}{\bibfnamefont{M.}~\bibnamefont{Scholer}},
  \bibinfo{journal}{Physics of Plasmas} \textbf{\bibinfo{volume}{11}}
  (\bibinfo{year}{2004}).

\bibitem[{\citenamefont{Loureiro et~al.}(2013)\citenamefont{Loureiro,
  Schekochihin, and Zocco}}]{loureiro:2013}
\bibinfo{author}{\bibfnamefont{N.~F.} \bibnamefont{Loureiro}},
  \bibinfo{author}{\bibfnamefont{A.~A.} \bibnamefont{Schekochihin}},
  \bibnamefont{and} \bibinfo{author}{\bibfnamefont{A.}~\bibnamefont{Zocco}},
  \bibinfo{journal}{Phys. Rev. Lett.} \textbf{\bibinfo{volume}{111}},
  \bibinfo{pages}{025002} (\bibinfo{year}{2013}),
  \urlprefix\url{http://link.aps.org/doi/10.1103/PhysRevLett.111.025002}.

\bibitem[{\citenamefont{Eastwood et~al.}(2013)\citenamefont{Eastwood, Phan,
  Drake, Shay, Borg, Lavraud, and Taylor}}]{eastwood:2013}
\bibinfo{author}{\bibfnamefont{J.~P.} \bibnamefont{Eastwood}},
  \bibinfo{author}{\bibfnamefont{T.~D.} \bibnamefont{Phan}},
  \bibinfo{author}{\bibfnamefont{J.~F.} \bibnamefont{Drake}},
  \bibinfo{author}{\bibfnamefont{M.~A.} \bibnamefont{Shay}},
  \bibinfo{author}{\bibfnamefont{A.~L.} \bibnamefont{Borg}},
  \bibinfo{author}{\bibfnamefont{B.}~\bibnamefont{Lavraud}}, \bibnamefont{and}
  \bibinfo{author}{\bibfnamefont{M.~G. G.~T.} \bibnamefont{Taylor}},
  \bibinfo{journal}{Phys. Rev. Lett.} \textbf{\bibinfo{volume}{110}},
  \bibinfo{pages}{225001} (\bibinfo{year}{2013}),
  \urlprefix\url{http://link.aps.org/doi/10.1103/PhysRevLett.110.225001}.

\bibitem[{\citenamefont{Yamada et~al.}(2015)\citenamefont{Yamada, Yoo,
  Jara-Almonte, Daughton, Ji, Kulsrud, and Myers}}]{yamada:2015}
\bibinfo{author}{\bibfnamefont{M.}~\bibnamefont{Yamada}},
  \bibinfo{author}{\bibfnamefont{J.}~\bibnamefont{Yoo}},
  \bibinfo{author}{\bibfnamefont{J.}~\bibnamefont{Jara-Almonte}},
  \bibinfo{author}{\bibfnamefont{W.}~\bibnamefont{Daughton}},
  \bibinfo{author}{\bibfnamefont{H.}~\bibnamefont{Ji}},
  \bibinfo{author}{\bibfnamefont{R.~M.} \bibnamefont{Kulsrud}},
  \bibnamefont{and} \bibinfo{author}{\bibfnamefont{C.~E.} \bibnamefont{Myers}},
  \bibinfo{journal}{Physics of Plasmas} \textbf{\bibinfo{volume}{22}},
  \bibinfo{eid}{056501} (\bibinfo{year}{2015}),
  \urlprefix\url{http://scitation.aip.org/content/aip/journal/pop/22/5/10.1063%
/1.4920960}.

\bibitem[{\citenamefont{Burch et~al.}(2014)\citenamefont{Burch, Moore, Torbert,
  and Giles}}]{burch:2014}
\bibinfo{author}{\bibfnamefont{J.}~\bibnamefont{Burch}},
  \bibinfo{author}{\bibfnamefont{T.}~\bibnamefont{Moore}},
  \bibinfo{author}{\bibfnamefont{R.}~\bibnamefont{Torbert}}, \bibnamefont{and}
  \bibinfo{author}{\bibfnamefont{B.}~\bibnamefont{Giles}},
  \bibinfo{journal}{Space Science Reviews} pp. \bibinfo{pages}{1--17}
  (\bibinfo{year}{2014}).

\bibitem[{\citenamefont{Hesse et~al.}(1999)\citenamefont{Hesse, Schindler,
  Birn, and Kuznetsova}}]{hesse:1999}
\bibinfo{author}{\bibfnamefont{M.}~\bibnamefont{Hesse}},
  \bibinfo{author}{\bibfnamefont{K.}~\bibnamefont{Schindler}},
  \bibinfo{author}{\bibfnamefont{J.}~\bibnamefont{Birn}}, \bibnamefont{and}
  \bibinfo{author}{\bibfnamefont{M.}~\bibnamefont{Kuznetsova}},
  \bibinfo{journal}{Physics of Plasmas} \textbf{\bibinfo{volume}{6}},
  \bibinfo{pages}{1781} (\bibinfo{year}{1999}),
  \urlprefix\url{http://link.aip.org/link/?PHP/6/1781/1}.

\bibitem[{\citenamefont{Ricci et~al.}(2003)\citenamefont{Ricci, Lapenta, and
  Brackbill}}]{ricci:2003}
\bibinfo{author}{\bibfnamefont{P.}~\bibnamefont{Ricci}},
  \bibinfo{author}{\bibfnamefont{G.}~\bibnamefont{Lapenta}}, \bibnamefont{and}
  \bibinfo{author}{\bibfnamefont{J.~U.} \bibnamefont{Brackbill}},
  \bibinfo{journal}{Physics of Plasmas} \textbf{\bibinfo{volume}{10}}
  (\bibinfo{year}{2003}).

\bibitem[{\citenamefont{Le et~al.}(2010{\natexlab{a}})\citenamefont{Le, Egedal,
  Daughton, Drake, Fox, and Katz}}]{le:2010grl}
\bibinfo{author}{\bibfnamefont{A.}~\bibnamefont{Le}},
  \bibinfo{author}{\bibfnamefont{J.}~\bibnamefont{Egedal}},
  \bibinfo{author}{\bibfnamefont{W.}~\bibnamefont{Daughton}},
  \bibinfo{author}{\bibfnamefont{J.~F.} \bibnamefont{Drake}},
  \bibinfo{author}{\bibfnamefont{W.}~\bibnamefont{Fox}}, \bibnamefont{and}
  \bibinfo{author}{\bibfnamefont{N.}~\bibnamefont{Katz}},
  \bibinfo{journal}{Geophys. Res. Lett.} \textbf{\bibinfo{volume}{37}},
  \bibinfo{pages}{L03106} (\bibinfo{year}{2010}{\natexlab{a}}), ISSN
  \bibinfo{issn}{0094-8276}.

\bibitem[{\citenamefont{Egedal et~al.}(2008)\citenamefont{Egedal, Fox, Katz,
  Porkolab, Oieroset, Lin, Daughton, and Drake}}]{egedal:2008jgr}
\bibinfo{author}{\bibfnamefont{J.}~\bibnamefont{Egedal}},
  \bibinfo{author}{\bibfnamefont{W.}~\bibnamefont{Fox}},
  \bibinfo{author}{\bibfnamefont{N.}~\bibnamefont{Katz}},
  \bibinfo{author}{\bibfnamefont{M.}~\bibnamefont{Porkolab}},
  \bibinfo{author}{\bibfnamefont{M.}~\bibnamefont{Oieroset}},
  \bibinfo{author}{\bibfnamefont{R.~P.} \bibnamefont{Lin}},
  \bibinfo{author}{\bibfnamefont{W.}~\bibnamefont{Daughton}}, \bibnamefont{and}
  \bibinfo{author}{\bibfnamefont{J.~F.} \bibnamefont{Drake}},
  \bibinfo{journal}{J. Geophys. Res.} \textbf{\bibinfo{volume}{113}},
  \bibinfo{pages}{A12207} (\bibinfo{year}{2008}), ISSN \bibinfo{issn}{A12207}.

\bibitem[{\citenamefont{Ng et~al.}(2011)\citenamefont{Ng, Egedal, Le, Daughton,
  and Chen}}]{ng:2011}
\bibinfo{author}{\bibfnamefont{J.}~\bibnamefont{Ng}},
  \bibinfo{author}{\bibfnamefont{J.}~\bibnamefont{Egedal}},
  \bibinfo{author}{\bibfnamefont{A.}~\bibnamefont{Le}},
  \bibinfo{author}{\bibfnamefont{W.}~\bibnamefont{Daughton}}, \bibnamefont{and}
  \bibinfo{author}{\bibfnamefont{L.~J.} \bibnamefont{Chen}},
  \bibinfo{journal}{Phys. Rev. Lett.} \textbf{\bibinfo{volume}{106}}
  (\bibinfo{year}{2011}), ISSN \bibinfo{issn}{0031-9007}.

\bibitem[{\citenamefont{Shuster et~al.}(2015)\citenamefont{Shuster, Chen,
  Hesse, Argall, Daughton, Torbert, and Bessho}}]{shuster:2015}
\bibinfo{author}{\bibfnamefont{J.~R.} \bibnamefont{Shuster}},
  \bibinfo{author}{\bibfnamefont{L.-J.} \bibnamefont{Chen}},
  \bibinfo{author}{\bibfnamefont{M.}~\bibnamefont{Hesse}},
  \bibinfo{author}{\bibfnamefont{M.~R.} \bibnamefont{Argall}},
  \bibinfo{author}{\bibfnamefont{W.}~\bibnamefont{Daughton}},
  \bibinfo{author}{\bibfnamefont{R.~B.} \bibnamefont{Torbert}},
  \bibnamefont{and} \bibinfo{author}{\bibfnamefont{N.}~\bibnamefont{Bessho}},
  \bibinfo{journal}{Geophysical Research Letters}
  \textbf{\bibinfo{volume}{42}}, \bibinfo{pages}{2586} (\bibinfo{year}{2015}),
  ISSN \bibinfo{issn}{1944-8007}, \bibinfo{note}{2015GL063601},
  \urlprefix\url{http://dx.doi.org/10.1002/2015GL063601}.

\bibitem[{\citenamefont{Phan et~al.}(2013)\citenamefont{Phan, Shay, Gosling,
  Fujimoto, Drake, Paschmann, Oieroset, Eastwood, and
  Angelopoulos}}]{phan:2013}
\bibinfo{author}{\bibfnamefont{T.~D.} \bibnamefont{Phan}},
  \bibinfo{author}{\bibfnamefont{M.~A.} \bibnamefont{Shay}},
  \bibinfo{author}{\bibfnamefont{J.~T.} \bibnamefont{Gosling}},
  \bibinfo{author}{\bibfnamefont{M.}~\bibnamefont{Fujimoto}},
  \bibinfo{author}{\bibfnamefont{J.~F.} \bibnamefont{Drake}},
  \bibinfo{author}{\bibfnamefont{G.}~\bibnamefont{Paschmann}},
  \bibinfo{author}{\bibfnamefont{M.}~\bibnamefont{Oieroset}},
  \bibinfo{author}{\bibfnamefont{J.~P.} \bibnamefont{Eastwood}},
  \bibnamefont{and}
  \bibinfo{author}{\bibfnamefont{V.}~\bibnamefont{Angelopoulos}},
  \bibinfo{journal}{Geophysical Research Letters}
  \textbf{\bibinfo{volume}{40}}, \bibinfo{pages}{4475} (\bibinfo{year}{2013}),
  ISSN \bibinfo{issn}{1944-8007},
  \urlprefix\url{http://dx.doi.org/10.1002/grl.50917}.

\bibitem[{\citenamefont{Shay et~al.}(2014)\citenamefont{Shay, Haggerty, Phan,
  Drake, Cassak, Wu, Oieroset, Swisdak, and Malakit}}]{shay:2014}
\bibinfo{author}{\bibfnamefont{M.~A.} \bibnamefont{Shay}},
  \bibinfo{author}{\bibfnamefont{C.~C.} \bibnamefont{Haggerty}},
  \bibinfo{author}{\bibfnamefont{T.~D.} \bibnamefont{Phan}},
  \bibinfo{author}{\bibfnamefont{J.~F.} \bibnamefont{Drake}},
  \bibinfo{author}{\bibfnamefont{P.~A.} \bibnamefont{Cassak}},
  \bibinfo{author}{\bibfnamefont{P.}~\bibnamefont{Wu}},
  \bibinfo{author}{\bibfnamefont{M.}~\bibnamefont{Oieroset}},
  \bibinfo{author}{\bibfnamefont{M.}~\bibnamefont{Swisdak}}, \bibnamefont{and}
  \bibinfo{author}{\bibfnamefont{K.}~\bibnamefont{Malakit}},
  \bibinfo{journal}{Physics of Plasmas} \textbf{\bibinfo{volume}{21}},
  \bibinfo{eid}{122902} (\bibinfo{year}{2014}),
  \urlprefix\url{http://scitation.aip.org/content/aip/journal/pop/21/12/10.106%
3/1.4904203}.

\bibitem[{\citenamefont{Le et~al.}(2014)\citenamefont{Le, Egedal, Ng,
  Karimabadi, Scudder, Roytershteyn, Daughton, and Liu}}]{le:2014}
\bibinfo{author}{\bibfnamefont{A.}~\bibnamefont{Le}},
  \bibinfo{author}{\bibfnamefont{J.}~\bibnamefont{Egedal}},
  \bibinfo{author}{\bibfnamefont{J.}~\bibnamefont{Ng}},
  \bibinfo{author}{\bibfnamefont{H.}~\bibnamefont{Karimabadi}},
  \bibinfo{author}{\bibfnamefont{J.}~\bibnamefont{Scudder}},
  \bibinfo{author}{\bibfnamefont{V.}~\bibnamefont{Roytershteyn}},
  \bibinfo{author}{\bibfnamefont{W.}~\bibnamefont{Daughton}}, \bibnamefont{and}
  \bibinfo{author}{\bibfnamefont{Y.-H.} \bibnamefont{Liu}},
  \bibinfo{journal}{Physics of Plasmas (1994-present)}
  \textbf{\bibinfo{volume}{21}}, \bibinfo{pages}{012103}
  (\bibinfo{year}{2014}).

\bibitem[{\citenamefont{Egedal et~al.}(2015)\citenamefont{Egedal, Daughton, Le,
  and Borg}}]{egedal:2015}
\bibinfo{author}{\bibfnamefont{J.}~\bibnamefont{Egedal}},
  \bibinfo{author}{\bibfnamefont{W.}~\bibnamefont{Daughton}},
  \bibinfo{author}{\bibfnamefont{A.}~\bibnamefont{Le}}, \bibnamefont{and}
  \bibinfo{author}{\bibfnamefont{A.~L.} \bibnamefont{Borg}},
  \bibinfo{journal}{Physics of Plasmas} \textbf{\bibinfo{volume}{22}},
  \bibinfo{eid}{101208} (\bibinfo{year}{2015}),
  \urlprefix\url{http://scitation.aip.org/content/aip/journal/pop/22/10/10.106%
3/1.4933055}.

\bibitem[{\citenamefont{Dahlin et~al.}(2014)\citenamefont{Dahlin, Drake, and
  Swisdak}}]{dahlin:2014}
\bibinfo{author}{\bibfnamefont{J.}~\bibnamefont{Dahlin}},
  \bibinfo{author}{\bibfnamefont{J.}~\bibnamefont{Drake}}, \bibnamefont{and}
  \bibinfo{author}{\bibfnamefont{M.}~\bibnamefont{Swisdak}},
  \bibinfo{journal}{Physics of Plasmas (1994-present)}
  \textbf{\bibinfo{volume}{21}}, \bibinfo{pages}{092304}
  (\bibinfo{year}{2014}).

\bibitem[{\citenamefont{Bowers et~al.}(2008)\citenamefont{Bowers, Albright,
  Yin, Bergen, and Kwan}}]{bowers:2008}
\bibinfo{author}{\bibfnamefont{K.~J.} \bibnamefont{Bowers}},
  \bibinfo{author}{\bibfnamefont{B.~J.} \bibnamefont{Albright}},
  \bibinfo{author}{\bibfnamefont{L.}~\bibnamefont{Yin}},
  \bibinfo{author}{\bibfnamefont{B.}~\bibnamefont{Bergen}}, \bibnamefont{and}
  \bibinfo{author}{\bibfnamefont{T.~J.~T.} \bibnamefont{Kwan}},
  \bibinfo{journal}{Physics of Plasmas} \textbf{\bibinfo{volume}{15}},
  \bibinfo{eid}{055703} (pages~\bibinfo{numpages}{7}) (\bibinfo{year}{2008}),
  \urlprefix\url{http://link.aip.org/link/?PHP/15/055703/1}.

\bibitem[{\citenamefont{Daughton et~al.}(2006)\citenamefont{Daughton, Scudder,
  and Karimabadi}}]{daughton:2006}
\bibinfo{author}{\bibfnamefont{W.}~\bibnamefont{Daughton}},
  \bibinfo{author}{\bibfnamefont{J.}~\bibnamefont{Scudder}}, \bibnamefont{and}
  \bibinfo{author}{\bibfnamefont{H.}~\bibnamefont{Karimabadi}},
  \bibinfo{journal}{Phys. Plasmas} \textbf{\bibinfo{volume}{13}},
  \bibinfo{pages}{072101} (\bibinfo{year}{2006}), ISSN
  \bibinfo{issn}{1070-664X}.

\bibitem[{\citenamefont{Le et~al.}({2009})\citenamefont{Le, Egedal, Daughton,
  Fox, and Katz}}]{le:2009}
\bibinfo{author}{\bibfnamefont{A.}~\bibnamefont{Le}},
  \bibinfo{author}{\bibfnamefont{J.}~\bibnamefont{Egedal}},
  \bibinfo{author}{\bibfnamefont{W.}~\bibnamefont{Daughton}},
  \bibinfo{author}{\bibfnamefont{W.}~\bibnamefont{Fox}}, \bibnamefont{and}
  \bibinfo{author}{\bibfnamefont{N.}~\bibnamefont{Katz}},
  \bibinfo{journal}{Phys. Rev. Lett.} \textbf{\bibinfo{volume}{{102}}},
  \bibinfo{pages}{{085001}} (\bibinfo{year}{{2009}}), ISSN
  \bibinfo{issn}{{0031-9007}}.

\bibitem[{\citenamefont{Le et~al.}(2013)\citenamefont{Le, Egedal, Ohia,
  Daughton, Karimabadi, and Lukin}}]{le:2013}
\bibinfo{author}{\bibfnamefont{A.}~\bibnamefont{Le}},
  \bibinfo{author}{\bibfnamefont{J.}~\bibnamefont{Egedal}},
  \bibinfo{author}{\bibfnamefont{O.}~\bibnamefont{Ohia}},
  \bibinfo{author}{\bibfnamefont{W.}~\bibnamefont{Daughton}},
  \bibinfo{author}{\bibfnamefont{H.}~\bibnamefont{Karimabadi}},
  \bibnamefont{and} \bibinfo{author}{\bibfnamefont{V.~S.} \bibnamefont{Lukin}},
  \bibinfo{journal}{Phys. Rev. Lett.} \textbf{\bibinfo{volume}{110}},
  \bibinfo{pages}{135004} (\bibinfo{year}{2013}),
  \urlprefix\url{http://link.aps.org/doi/10.1103/PhysRevLett.110.135004}.

\bibitem[{\citenamefont{Chen et~al.}(2012)\citenamefont{Chen, Daughton,
  Bhattacharjee, Torbert, Roytershteyn, and Bessho}}]{chen:2012}
\bibinfo{author}{\bibfnamefont{L.-J.} \bibnamefont{Chen}},
  \bibinfo{author}{\bibfnamefont{W.}~\bibnamefont{Daughton}},
  \bibinfo{author}{\bibfnamefont{A.}~\bibnamefont{Bhattacharjee}},
  \bibinfo{author}{\bibfnamefont{R.~B.} \bibnamefont{Torbert}},
  \bibinfo{author}{\bibfnamefont{V.}~\bibnamefont{Roytershteyn}},
  \bibnamefont{and} \bibinfo{author}{\bibfnamefont{N.}~\bibnamefont{Bessho}},
  \bibinfo{journal}{Physics of Plasmas} \textbf{\bibinfo{volume}{19}},
  \bibinfo{eid}{112902} (\bibinfo{year}{2012}),
  \urlprefix\url{http://scitation.aip.org/content/aip/journal/pop/19/11/10.106%
3/1.4767645}.

\bibitem[{\citenamefont{Che et~al.}(2010)\citenamefont{Che, Drake, Swisdak, and
  Yoon}}]{che:2010}
\bibinfo{author}{\bibfnamefont{H.}~\bibnamefont{Che}},
  \bibinfo{author}{\bibfnamefont{J.~F.} \bibnamefont{Drake}},
  \bibinfo{author}{\bibfnamefont{M.}~\bibnamefont{Swisdak}}, \bibnamefont{and}
  \bibinfo{author}{\bibfnamefont{P.~H.} \bibnamefont{Yoon}},
  \bibinfo{journal}{Geophysical Research Letters}
  \textbf{\bibinfo{volume}{37}}, \bibinfo{pages}{n/a} (\bibinfo{year}{2010}),
  ISSN \bibinfo{issn}{1944-8007}, \bibinfo{note}{l11105},
  \urlprefix\url{http://dx.doi.org/10.1029/2010GL043608}.

\bibitem[{\citenamefont{Roytershteyn et~al.}(2012)\citenamefont{Roytershteyn,
  Daughton, Karimabadi, and Mozer}}]{roytershteyn:2012}
\bibinfo{author}{\bibfnamefont{V.}~\bibnamefont{Roytershteyn}},
  \bibinfo{author}{\bibfnamefont{W.}~\bibnamefont{Daughton}},
  \bibinfo{author}{\bibfnamefont{H.}~\bibnamefont{Karimabadi}},
  \bibnamefont{and} \bibinfo{author}{\bibfnamefont{F.~S.} \bibnamefont{Mozer}},
  \bibinfo{journal}{Phys. Rev. Lett.} \textbf{\bibinfo{volume}{108}},
  \bibinfo{pages}{185001} (\bibinfo{year}{2012}),
  \urlprefix\url{http://link.aps.org/doi/10.1103/PhysRevLett.108.185001}.

\bibitem[{\citenamefont{Egedal et~al.}({2013})\citenamefont{Egedal, Le, and
  Daughton}}]{egedal:2013pop}
\bibinfo{author}{\bibfnamefont{J.}~\bibnamefont{Egedal}},
  \bibinfo{author}{\bibfnamefont{A.}~\bibnamefont{Le}}, \bibnamefont{and}
  \bibinfo{author}{\bibfnamefont{W.}~\bibnamefont{Daughton}},
  \bibinfo{journal}{Phys. Plasmas} \textbf{\bibinfo{volume}{{20}}}
  (\bibinfo{year}{{2013}}), ISSN \bibinfo{issn}{{1070-664X}}.

\bibitem[{\citenamefont{Speiser}(1965)}]{speiser:1965}
\bibinfo{author}{\bibfnamefont{T.}~\bibnamefont{Speiser}}, \bibinfo{journal}{J.
  Geophys. Res.} \textbf{\bibinfo{volume}{70}}, \bibinfo{pages}{4219}
  (\bibinfo{year}{1965}).

\bibitem[{\citenamefont{Bessho et~al.}(2014)\citenamefont{Bessho, Chen,
  Shuster, and Wang}}]{bessho:2014}
\bibinfo{author}{\bibfnamefont{N.}~\bibnamefont{Bessho}},
  \bibinfo{author}{\bibfnamefont{L.-J.} \bibnamefont{Chen}},
  \bibinfo{author}{\bibfnamefont{J.~R.} \bibnamefont{Shuster}},
  \bibnamefont{and} \bibinfo{author}{\bibfnamefont{S.}~\bibnamefont{Wang}},
  \bibinfo{journal}{Geophysical Research Letters}
  \textbf{\bibinfo{volume}{41}}, \bibinfo{pages}{8688} (\bibinfo{year}{2014}),
  ISSN \bibinfo{issn}{1944-8007},
  \urlprefix\url{http://dx.doi.org/10.1002/2014GL062034}.

\bibitem[{\citenamefont{Buchner and Zelenyi}(1989)}]{buchner:1989}
\bibinfo{author}{\bibfnamefont{J.}~\bibnamefont{Buchner}} \bibnamefont{and}
  \bibinfo{author}{\bibfnamefont{L.}~\bibnamefont{Zelenyi}},
  \bibinfo{journal}{J. Geophys. Res.} \textbf{\bibinfo{volume}{94}},
  \bibinfo{pages}{11821} (\bibinfo{year}{1989}).

\bibitem[{\citenamefont{Chen et~al.}(2008)\citenamefont{Chen, Bhattacharjee,
  Puhl-Quinn, Yang, Bessho, Imada, Muehlbachler, Daly, Lefebvre, Khotyaintsev
  et~al.}}]{chen:2008}
\bibinfo{author}{\bibfnamefont{L.~J.} \bibnamefont{Chen}},
  \bibinfo{author}{\bibfnamefont{A.}~\bibnamefont{Bhattacharjee}},
  \bibinfo{author}{\bibfnamefont{P.~A.} \bibnamefont{Puhl-Quinn}},
  \bibinfo{author}{\bibfnamefont{H.}~\bibnamefont{Yang}},
  \bibinfo{author}{\bibfnamefont{N.}~\bibnamefont{Bessho}},
  \bibinfo{author}{\bibfnamefont{S.}~\bibnamefont{Imada}},
  \bibinfo{author}{\bibfnamefont{S.}~\bibnamefont{Muehlbachler}},
  \bibinfo{author}{\bibfnamefont{P.~W.} \bibnamefont{Daly}},
  \bibinfo{author}{\bibfnamefont{B.}~\bibnamefont{Lefebvre}},
  \bibinfo{author}{\bibfnamefont{Y.}~\bibnamefont{Khotyaintsev}},
  \bibnamefont{et~al.}, \bibinfo{journal}{Nature Physics}
  \textbf{\bibinfo{volume}{4}}, \bibinfo{pages}{19} (\bibinfo{year}{2008}),
  ISSN \bibinfo{issn}{1745-2473}.

\bibitem[{\citenamefont{Le et~al.}(2010{\natexlab{b}})\citenamefont{Le, Egedal,
  Fox, Katz, Vrublevskis, Daughton, and Drake}}]{le:2010pop}
\bibinfo{author}{\bibfnamefont{A.}~\bibnamefont{Le}},
  \bibinfo{author}{\bibfnamefont{J.}~\bibnamefont{Egedal}},
  \bibinfo{author}{\bibfnamefont{W.}~\bibnamefont{Fox}},
  \bibinfo{author}{\bibfnamefont{N.}~\bibnamefont{Katz}},
  \bibinfo{author}{\bibfnamefont{A.}~\bibnamefont{Vrublevskis}},
  \bibinfo{author}{\bibfnamefont{W.}~\bibnamefont{Daughton}}, \bibnamefont{and}
  \bibinfo{author}{\bibfnamefont{J.~F.} \bibnamefont{Drake}},
  \bibinfo{journal}{Phys. Plasmas} \textbf{\bibinfo{volume}{17}},
  \bibinfo{pages}{055703} (\bibinfo{year}{2010}{\natexlab{b}}), ISSN
  \bibinfo{issn}{1070-664X}, \bibinfo{note}{51st Annual Meeting of the Division
  of Plasma Physics of the American Physical Society, Atlanta, GA, NOV 02-06,
  2009}.

\bibitem[{\citenamefont{Nakamura et~al.}(2016)\citenamefont{Nakamura, Nakamura,
  and Haseagwa}}]{nakamura:2016}
\bibinfo{author}{\bibfnamefont{T.}~\bibnamefont{Nakamura}},
  \bibinfo{author}{\bibfnamefont{R.}~\bibnamefont{Nakamura}}, \bibnamefont{and}
  \bibinfo{author}{\bibfnamefont{H.}~\bibnamefont{Haseagwa}}, in
  \emph{\bibinfo{booktitle}{Annales Geophysicae}}
  (\bibinfo{organization}{Copernicus GmbH}, \bibinfo{year}{2016}),
  vol.~\bibinfo{volume}{34}, pp. \bibinfo{pages}{357--367}.

\bibitem[{\citenamefont{Hesse et~al.}(2008)\citenamefont{Hesse, Zenitani, and
  Klimas}}]{hesse:2008}
\bibinfo{author}{\bibfnamefont{M.}~\bibnamefont{Hesse}},
  \bibinfo{author}{\bibfnamefont{S.}~\bibnamefont{Zenitani}}, \bibnamefont{and}
  \bibinfo{author}{\bibfnamefont{A.}~\bibnamefont{Klimas}},
  \bibinfo{journal}{Phys. Plasmas} \textbf{\bibinfo{volume}{15}},
  \bibinfo{eid}{112102} (pages~\bibinfo{numpages}{5}) (\bibinfo{year}{2008}).

\bibitem[{\citenamefont{Divin et~al.}(2012)\citenamefont{Divin, Lapenta,
  Markidis, Semenov, Erkaev, Korovinskiy, and Biernat}}]{divin:2012}
\bibinfo{author}{\bibfnamefont{A.}~\bibnamefont{Divin}},
  \bibinfo{author}{\bibfnamefont{G.}~\bibnamefont{Lapenta}},
  \bibinfo{author}{\bibfnamefont{S.}~\bibnamefont{Markidis}},
  \bibinfo{author}{\bibfnamefont{V.~S.} \bibnamefont{Semenov}},
  \bibinfo{author}{\bibfnamefont{N.~V.} \bibnamefont{Erkaev}},
  \bibinfo{author}{\bibfnamefont{D.~B.} \bibnamefont{Korovinskiy}},
  \bibnamefont{and} \bibinfo{author}{\bibfnamefont{H.~K.}
  \bibnamefont{Biernat}}, \bibinfo{journal}{Journal of Geophysical Research:
  Space Physics} \textbf{\bibinfo{volume}{117}}, \bibinfo{pages}{n/a}
  (\bibinfo{year}{2012}), ISSN \bibinfo{issn}{2156-2202},
  \bibinfo{note}{a06217},
  \urlprefix\url{http://dx.doi.org/10.1029/2011JA017464}.

\bibitem[{\citenamefont{Wu et~al.}(2011)\citenamefont{Wu, Shay, Phan, Oieroset,
  and Oka}}]{wu:2011}
\bibinfo{author}{\bibfnamefont{P.}~\bibnamefont{Wu}},
  \bibinfo{author}{\bibfnamefont{M.~A.} \bibnamefont{Shay}},
  \bibinfo{author}{\bibfnamefont{T.~D.} \bibnamefont{Phan}},
  \bibinfo{author}{\bibfnamefont{M.}~\bibnamefont{Oieroset}}, \bibnamefont{and}
  \bibinfo{author}{\bibfnamefont{M.}~\bibnamefont{Oka}},
  \bibinfo{journal}{Physics of Plasmas} \textbf{\bibinfo{volume}{18}},
  \bibinfo{eid}{111204} (\bibinfo{year}{2011}),
  \urlprefix\url{http://scitation.aip.org/content/aip/journal/pop/18/11/10.106%
3/1.3641964}.

\bibitem[{\citenamefont{Phan}(2015)}]{phan:pc}
\bibinfo{author}{\bibfnamefont{T.}~\bibnamefont{Phan}} (\bibinfo{year}{2015}),
  \bibinfo{note}{private communication}.

\bibitem[{\citenamefont{Ohia et~al.}(2015)\citenamefont{Ohia, Egedal, Lukin,
  Daughton, and Le}}]{ohia:2015}
\bibinfo{author}{\bibfnamefont{O.}~\bibnamefont{Ohia}},
  \bibinfo{author}{\bibfnamefont{J.}~\bibnamefont{Egedal}},
  \bibinfo{author}{\bibfnamefont{V.~S.} \bibnamefont{Lukin}},
  \bibinfo{author}{\bibfnamefont{W.}~\bibnamefont{Daughton}}, \bibnamefont{and}
  \bibinfo{author}{\bibfnamefont{A.}~\bibnamefont{Le}},
  \bibinfo{journal}{Geophysical Research Letters}
  \textbf{\bibinfo{volume}{42}}, \bibinfo{pages}{10,549}
  (\bibinfo{year}{2015}), ISSN \bibinfo{issn}{1944-8007},
  \bibinfo{note}{2015GL067117},
  \urlprefix\url{http://dx.doi.org/10.1002/2015GL067117}.

\end{thebibliography}

\end{document}